\documentstyle[12pt]{article}
\begin{document} 
\global\parskip 6pt
\newcommand{\be}{\begin{equation}}
\newcommand{\ee}{\end{equation}}
\newcommand{\bea}{\begin{eqnarray}}
\newcommand{\eea}{\end{eqnarray}}
\newcommand{\non}{\nonumber}

\begin{titlepage}
\hfill{hep-th/9812206}
\vspace*{1cm}
\begin{center}
{\Large\bf Geometrical Finiteness, Holography,}\\
\vspace*{.2cm}
{\Large \bf and the BTZ Black Hole}\\
\vspace*{2cm}
Danny Birmingham\footnote{E-mail: dannyb@pop3.ucd.ie}\\
\vspace*{.5cm}
{\em Department of Mathematical Physics\\
University College Dublin\\
Belfield, Dublin 4, Ireland}\\
\vspace*{1cm}
Conall Kennedy, Siddhartha Sen, Andy Wilkins\footnote{E-mail:
conall@maths.tcd.ie, sen@maths.tcd.ie, andyw@maths.tcd.ie}\\
\vspace*{.5cm}
{\em School of Mathematics\\
Trinity College Dublin\\
Dublin 2, Ireland}\\
\vspace{2cm}

\begin{abstract}

We show how a theorem of Sullivan provides a precise mathematical
statement of a 3d holographic principle, that is, the hyperbolic
structure of a certain class of 3d manifolds is completely
determined in terms of the corresponding Teichm\"{u}ller space of the
boundary. We explore the consequences of this theorem
in the context of the Euclidean BTZ black hole in three dimensions.
\end{abstract}
\vspace{1cm}
December 1998
\end{center}
\end{titlepage}

The holographic principle as formulated by Maldacena
\cite{Malda} conjectures a dynamical
correspondence between string theory on anti-de Sitter backgrounds,
and conformal field theory defined on the boundary of anti-de Sitter space.
The precise nature of this correspondence has been studied in
\cite{Poly}-\cite{Witten2}, and various pieces of evidence in favour
of the conjecture have been established.

In $2+1$ dimensions, the Maldacena conjecture can be investigated
more thoroughly,
since details are known about both sides of the
adS/CFT correspondence \cite{MS,Mart}.
Solutions to Einstein's equations with negative cosmological constant
form the basic ingredients in the correspondence, and the
BTZ black hole \cite{BTZ1,BTZ2} provides an example of such a solution
in $2+1$ dimensions.
In \cite{BH}, it was shown that the asymptotic symmetry algebra
of $adS_{3}$ consists of two copies of the Virasoro algebra.
The mass and angular momentum of the BTZ black hole can then be written
in terms of the Virasoro generators $L_{0} \pm \overline{L}_{0}$.
It was observed in \cite{Strom,BSS} that the Cardy formula
could be used to give precise agreement between the Bekenstein-Hawking
entropy and the entropy of the asymptotic conformal field theory.
Further aspects of the entropy issue within this context have been
considered in \cite{Carlip1}. In particular, the question arises as to the
location of the microstates which contribute to the entropy.
In \cite{Strom,BSS}, the entropy is accounted for by
the asymptotic symmetry algebra, while in \cite{Carlip2}
an attempt is made to understand the entropy in terms of near-horizon
microstates.

In this paper, we draw attention to the fact that there
is a precise notion of holography in the kinematical sense.
This depends on a theorem of Sullivan \cite{Sull},
see also \cite{McM1}, which states that the inequivalent hyperbolic
structures of a three-dimensional geometrically finite Kleinian
manifold (a term to be defined) are parametrized by the Teichm\"{u}ller
space of the boundary.
Thus, the theorem of Sullivan provides a precise mathematical statement
of a 3d holographic principle.
Our aim here is to explore the consequences of this theorem in the context
of the Euclidean BTZ black hole, hereafter referred to as the
BTZ black hole.
We first recall that the BTZ black hole
is obtained as a quotient  of hyperbolic
$3$-space $H^{3}$ (a space of constant negative curvature)
by a discrete subgroup $\Gamma$ of
isometries of $H^{3}$ (a Kleinian group) \cite{BTZ2}.
We then show that the resulting
Kleinian manifold $M$, which necessarily has constant
negative curvature,  is geometrically finite with the topology
of a solid torus; the boundary $\partial M$ is
thus a $2$-torus.
A consequence of the holographic theorem of Sullivan is that the
hyperbolic geometry
of the BTZ black hole is completely specified
by the Teichm\"{u}ller class of the genus one boundary. Since
the Teichm\"{u}ller class is determined by a single complex parameter,
the theorem can also be viewed as providing a `No Hair' theorem
for the BTZ black hole. In other words, the BTZ black hole
can be parametrized by at most two
real parameters; these parameters may be taken to be the mass
and angular momentum, for example.

A further consequence of this holographic theorem relates
to the entropy discussion. Since the Bekenstein-Hawking entropy
is a geometrical quantity, it is determined once the hyperbolic
geometry is fixed, which is in turn fixed by the Teichm\"{u}ller
parameters. Thus, the entropy is completely determined
in terms of the boundary data, as
advocated in \cite{Strom,BSS}.

While the dynamical conjecture of Maldacena remains to be proved,
we stress
that the kinematical result of Sullivan is a precise mathematical theorem.
As such, it provides an improved description of the basic
geometry of the BTZ black hole. It is vital to have a clear
picture of the underlying mathematical structure, and the theorem
of Sullivan contributes to this understanding in
a precise way.

We begin by recalling that the BTZ metric in Schwarzschild
coordinates is given by \cite{Carlip3,Carlip4}
\bea
ds_{E}^{2} = \left(-M + \frac{r^{2}}{l^{2}} -
\frac{J_{E}^{2}}{4r^{2}}\right)
d\tau_{E}^{2} + \left(-M + \frac{r^{2}}{l^{2}} -
\frac{J_{E}^{2}}{4r^{2}}\right)^{-1}
dr^{2} + r^{2} \left( d\phi - \frac{J_{E}}{2r^{2}}d\tau_{E}\right)^{2},
\eea
where
$M$ and $J_{E}$ are the mass and angular momentum parameters of the black
hole.

It is known that the BTZ black hole is obtained
by performing a quotient of hyperbolic $3$-space $H^{3}$
by a discrete finitely generated group of isometries
of $H^{3}$ \cite{BTZ2}.
The precise form of the identifications can be seen
most easily by writing the metric for $H^{3}$
in the upper-half space coordinates $(x,y,z)$ defined by
\cite{Carlip3,Carlip4}
\bea
x &=& \left(\frac{r^{2} - r_{+}^{2}}{r^{2} - r_{-}^{2}}\right)^{1/2}\;
\cos\left( \frac{r_{+}}{l^{2}}\tau_{E} + \frac{\mid\! r_{-}\!
\mid}{l}\phi\right)\; \exp\left(\frac{r_{+}}{l}\phi
- \frac{\mid\! r_{-}
\!\mid}{l^{2}} \tau_{E}\right),\non\\
y &=& \left(\frac{r^{2} - r_{+}^{2}}{r^{2} - r_{-}^{2}}\right)^{1/2}\;
\sin\left( \frac{r_{+}}{l^{2}}\tau_{E} + \frac{\mid\! r_{-}\!
\mid}{l}\phi\right)\; \exp\left(\frac{r_{+}}{l}\phi - \frac{\mid\! r_{-}\!
\mid}{l^{2}} \tau_{E}\right),\non\\
z &=& \left(\frac{r_{+}^{2} - r_{-}^{2}}{r^{2} - r_{-}^{2}}\right)^{1/2}\;
\exp\left(\frac{r_{+}}{l}\phi - \frac{\mid\! r_{-}\!
\mid}{l^{2}} \tau_{E}\right),
\eea
where the parameters $r_{\pm}$ are defined by
\be
r_{\pm}^{2} = \frac{Ml^{2}}{2}\left(1 \pm \sqrt{1 +
\frac{J_{E}^{2}}{M^{2}l^{2}}}\;\right),
\ee
and
$\mid \! r_{-}\!\mid = ir_{-}$.
The metric becomes
\bea
ds_{E}^{2} = \frac{l^{2}}{z^{2}}(dx^{2} + dy^{2} + dz^{2}),\;\;
z>0.
\eea

As shown in \cite{Carlip3,Carlip4}, the periodicity in the Schwarzschild
coordinates $\phi$ is then implemented via the identifications
\bea
& &(x,y,z) \sim \label{Gamma}\\
& & e^{2\pi r_{+}/l}\left(
x \cos\left(\frac{2\pi\mid\! r_{-}\!\mid}{l}\right)
-y \sin\left(\frac{2\pi \mid\! r_{-}\!\mid}{l}\right),
x \sin\left(\frac{2\pi\mid\! r_{-}\!\mid}{l}\right)
+y \cos\left(\frac{2\pi \mid\! r_{-}\!\mid}{l}\right),
z\right).\non
\eea
Let $\Gamma$ be the discrete subgroup of isometries of $H^{3}$
generated by (\ref{Gamma}). Then clearly $\Gamma$ is a discrete
finitely
generated Kleinian group; in fact, $\Gamma$ is an abelian cyclic group.
Our aim now is to show that the resulting quotient manifold
$H^{3}/\Gamma$ is geometrically finite.
Relevant background material can be found for example in
\cite{Bowditch}-\cite{Kra}.

Let us consider the Poincar\'{e} ball model for hyperbolic $3$-space
\cite{Bowditch},
in which the surface at infinity is a $2$-sphere $S^{2}_{\infty}$.
This model is equivalent to the upper-half space model.
We denote by $\Gamma_{x}$ the orbit of any point of $H^{3}$ under
the action of $\Gamma$.
The {\em limit set} of $\Gamma$ is defined as
\be
L_{\Gamma} = \overline{\Gamma}_{x}\cap S^{2}_{\infty}.
\ee
Thus, the limit set is
the intersection of the closure of $\Gamma_{x}$ with the sphere
at infinity,
for any point $x \in H^{3}$. One can show that the limit set
is independent of the choice of $x$, as we will demonstrate
explicitly in the case at hand.
Given the limit set $L_{\Gamma}$,
one now defines the {\em convex hull}
$H(L_{\Gamma})$ to be the smallest convex set in $H^{3}$
containing $L_{\Gamma}$.
We recall that a convex set in $H^{3}$ is one which contains all
geodesics joining any two points in the set.
The associated {\em convex core}
is obtained as a quotient
\be
C(\Gamma) = H(L_{\Gamma})/\Gamma.
\ee
Thus, the convex core is the quotient by $\Gamma$ of the smallest
convex set in $H^{3}$ containing all geodesics with both end points
in the limit set.

We see that the action of $\Gamma$ partitions $S^{2}_{\infty}$ into
the limit set $L_{\Gamma}$ and its complement $\Omega$, which is
called the {\em domain of discontinuity}.
The resulting Kleinian manifold
\be
N = (H^{3} \cup  \Omega)/\Gamma,
\ee
is then a $3$-manifold with a hyperbolic structure on its interior
and a complex structure on its boundary \cite{McM1,Bowditch}.
A hyperbolic $3$-manifold is said to be a {\em geometrically finite}
Kleinian manifold if
the volume of the convex core of $\Gamma$ is finite.
As shown in \cite{Bowditch}, there are several equivalent definitions
of geometrical finiteness.
The main theorem may now be stated as follows.
Let $M$ denote a topological $3$-manifold, and let $GF(M)$ denote
the space of geometrically finite hyperbolic $3$-manifolds $N$
which are homeomorphic to $M$. Thus, $GF(M)$ denotes the space of
realizations of $M$ by geometrically finite Kleinian manifolds $N$.
Then, we have the following theorem due to Sullivan \cite{Sull},
see also \cite{McM1}.

\noindent {\bf Theorem}. As long as $M$ admits at least one
hyperbolic realization,
there is a 1-1 correspondence between hyperbolic structures on $M$
and conformal structures on $\partial M$, i.e.,
\be
GF(M) \cong \mathrm{Teich}(\partial M),
\ee
where $Teich(\partial M)$ is the Teichm\"{u}ller space of $\partial M$.

Our first aim is to show that the BTZ black hole
is a geometrically finite Kleinian manifold. We shall then discuss the
consequences of this in light of the theorem stated above. To establish
the geometrical finiteness,
the first order of business is to determine the limit
set. The orbit of any point in $H^{3}$ under the action
of the BTZ group (\ref{Gamma}) is easily obtained in the upper-half space
model. In this model, the boundary at infinity is the $z=0$
plane and the point $z=\infty$. One notes that $z=\infty$ corresponds
to a single point
since the metric in the $x,y$ directions vanishes as
$z\rightarrow \infty$ \cite{Witten1}.

If we let $\gamma$ denote the identification defined by (\ref{Gamma}),
then
clearly the BTZ group is a cyclic Kleinian group with elements
\be
\Gamma = \{\gamma^{n}, n \in {\mathbf Z}\}.
\ee
Thus, the limit set
of $\Gamma$ consists of two points, the origin
and the point $z=\infty$, corresponding to $n\rightarrow
\pm \infty$. A simple geometrical picture of the orbit of any point
is obtained by noting that the orbit has a helical structure, whereby
points are rotated around the $z$-axis, as they are translated
upwards or downwards along the $z$-axis. Note also that the limit set
is independent of the chosen point, as required.
Such isometries are referred to as
loxodromic, with the $z$-axis called the loxodromic axis \cite{Bowditch}.

It is known that if the limit set of a Kleinian group $\Gamma$ is finite,
then it must contain zero, one, or two points \cite{Kra}.
Such Kleinian groups are
called elementary. Furthermore, if $\Gamma$ is elementary, then it
is geometrically finite \cite{Maskit}.
Based on the fact that the limit set of the BTZ group
consists of two points, we immediately conclude that
it is a geometrically finite group, and the BTZ
black hole is a geometrically finite Kleinian manifold.
One can also proceed to check the geometrical finiteness according
to the equivalent definitions \cite{Bowditch}.

Having established the geometrical finiteness
of the BTZ black hole, we wish
to determine the topology of the resulting Kleinian
manifold.
This can be seen by introducing spherical coordinates on
the upper-half space defined by \cite{Carlip3,Carlip4}
\bea
(x,y,z) = (R \cos \theta \cos \chi, R \sin \theta \cos \chi,
R \sin \chi),
\eea
with $\theta \in[0,2\pi], \chi \in[0,\pi/2]$.
The line element is then written as
\bea
ds_{E}^{2} = \frac{l^{2}}{\sin^{2}\chi}\left( \frac{dR^{2}}{R^{2}}
+ d\chi^{2} + \cos^{2}\chi \;d\theta^{2}\right),
\eea
and the  identifications (\ref{Gamma}) become
\bea
(R,\theta,\chi) \sim
\left(R\;e^{2 \pi r_{+}/l}, \theta + \frac{2 \pi \mid\!
r_{-}\!\mid}{l}, \chi\right).
\eea
A fundamental region is the space between the hemispheres $R=1$ and
$R= e^{2\pi r_{+}/l}$ with the inner and outer boundaries
identified along a radial line, followed by a $2\pi r_{+}/l$ rotation
about the $z$-axis. Topologically, the resulting manifold is
a solid torus \cite{Carlip3,Carlip4}. For $\chi \neq \pi/2$,
each slice of fixed $\chi$ is a $2$-torus with periodic coordinates
$\ln R$ and $\theta$.

Thus, we have shown that the BTZ black hole is a geometrically
finite Kleinian manifold with the topology of a solid torus.
We can
now explore some of the consequences of the theorem of Sullivan.

\noindent (i) The theorem allows us to declare that the BTZ
manifold is a holographic manifold, such that
the three-dimensional hyperbolic structures are in 1-1
correspondence with the
Teichm\"{u}ller parameters of the two-dimensional genus one
boundary.

\noindent (ii) Since the Teichm\"{u}ller space is parametrized by
two real parameters,
the theorem states that we have a `No Hair' theorem; namely, the BTZ
black hole can be parametrized by at most two parameters, thus
excluding the construction of a charged rotating generalization
as a geometrically finite Kleinian manifold.
Attempts at generalizations are reviewed in \cite{Carlip3}.

\noindent (iii) As the Bekenstein-Hawking entropy formula is a
geometrical
quantity, it is determined once the hyperbolic structure is fixed.
Hence, the Bekenstein-Hawking entropy is determined
by the Teichm\"{u}ller space on the boundary.
This is in agreement with the observation in \cite{Strom,BSS}
based on the asymptotic symmetry algebra of anti-de Sitter space.

\noindent (iv) In \cite{MS}, the action for
the BTZ black hole and $adS_{3}$ were written in
terms of a complex temperature parameter $\tau$
(equivalently, the mass and angular momentum) defined on
the boundary $2$-torus.
The resulting form of the action then suggested the existence
of an
$SL(2,Z)$ family of solutions whose boundary data $\tau$
is related by the associated modular transformations. We see that the
above theorem does indeed establish the existence of
this class of hyperbolic geometries. Two such geometries whose
Teichm\"{u}ller parameters are related by a modular transformation
are then equivalent as hyperbolic structures.
Furthermore, it was noted in \cite{MS} that there is a correspondence
between the $SL(2,\mathbf{C})$ isometry used to construct the BTZ
black hole, and the $SL(2,\mathbf{C})$ element used in the boundary
conformal field theory. This correspondence finds a precise
explanation in the theorem of Sullivan.

\noindent (v) The observation that the BTZ black hole is a geometrically
finite Kleinian manifold with genus one boundary suggests
that one might try to construct more general black holes
which have higher genus boundary. The theorem of Sullivan
again applies since the corresponding Kleinian manifold is
geometrically finite \cite{Maskit}, and such objects
would be parametrized
by $(3g-3)$ complex Teichm\"{u}ller parameters. They
would also play a role in the adS/CFT correspondence, where
one would be studying conformal field theory on higher genus
Riemann surfaces.

In conclusion, the above analysis gives an exact example of a three-
dimensional holographic manifold, and thus clarifies the basic geometry
of the BTZ black hole.

\noindent {\large \bf Acknowledgements}\\
D.B. and S.S. would like to thank Martin Bridgeman for a discussion
on the theorem of Sullivan. This work is part of a project
supported by Enterprise Ireland Basic Research Grant SC/98/741.

\end{document}